\documentclass[aps,prl,twocolumn,groupedaddress]{revtex4-2}
\usepackage{graphicx}
\usepackage[compact]{titlesec}
\usepackage{amsmath}
\usepackage{physics}
\usepackage{multirow}
\usepackage{braket}
\usepackage{hyperref}
\usepackage{caption}
\usepackage[font=singlespacing]{caption}
\usepackage{float}
\usepackage{amsmath}
\usepackage{wrapfig}
\usepackage{subcaption}

\begin{document}

\title{Direct confirmation of long-range magnetic order and evidence for multipoles in Ce\textsubscript{2}O\textsubscript{3}}

\author{Alexandra Cote}
\affiliation{Department of Physics and Materials Research Laboratory, University of Illinois at Urbana-Champaign, Urbana, Illinois 61801, USA}
\author{J. Eddie Slimak}
\affiliation{Department of Physics and Materials Research Laboratory, University of Illinois at Urbana-Champaign, Urbana, Illinois 61801, USA}
\author{Astha Sethi}
\affiliation{Department of Physics and Materials Research Laboratory, University of Illinois at Urbana-Champaign, Urbana, Illinois 61801, USA}
\author{Dalmau Reig-i-Plessis}
\affiliation{Department of Physics and Astronomy and Stewart Blusson Quantum Matter Institute, University of British Columbia, Vancouver, Canada V6T 1Z4}
\author{Qiang Zhang}
\affiliation{Neutron Scattering Division, Oak Ridge National Laboratory, Oak Ridge, Tennessee 37831, USA}
\author{Yang Zhao}
\affiliation{NIST Center for Neutron Research, National Institute of Standards and Technology, Gaithersburg, Maryland 20899, USA and Department of Materials Science and Engineering, University of Maryland, College Park, Maryland 20742, USA }
\author{Devashibhai Adroja}
\affiliation{ISIS Neutron and Muon Facility, STFC, Rutherford Appleton Laboratory, Chilton, Didcot Oxon, OX11 0QX, United Kingdom and
Highly Correlated Matter Research Group, Physics Department,
University of Johannesburg, Auckland Park 2006, South Africa}
\author{Gerald Morris}
\affiliation{TRIUMF Centre for Molecular and Materials Science, Vancouver, BC V6T 2A3, Canada}
\author{Taras Kolodiazhnyi}
\affiliation{National Institute for Materials Science, 1-1 Namiki, Tsukuba, Ibaraki, 305-0044, Japan}
\author{Alannah M. Hallas}
\affiliation{Department of Physics and Astronomy and Stewart Blusson Quantum Matter Institute, University of British Columbia, Vancouver, Canada V6T 1Z4}
\author{Jeffrey W. Lynn}
\affiliation{NIST Center for Neutron Research, National Institute of Standards and Technology, Gaithersburg, Maryland 20899, USA}
\author{S. Lance Cooper}
\affiliation{Department of Physics and Materials Research Laboratory, University of Illinois at Urbana-Champaign, Urbana, Illinois 61801, USA}
\author{Gregory J. MacDougall}
\affiliation{Department of Physics and Materials Research Laboratory, University of Illinois at Urbana-Champaign, Urbana, Illinois 61801, USA}

\date{\today}

\begin{abstract}
The sesquioxide, Ce\textsubscript{2}O\textsubscript{3}, has been a material of intense interest in recent years due to reports of an anomalous giant magnetodielectric effect and emergent mixed crystal field-phonon (vibronic) excitations below a putative antiferromagnetic transition at T\textsubscript{N} = 6.2 K. The claim of long-range magnetic order in this material is based on heat capacity and temperature-dependent susceptibility measurements; however, multiple neutron diffraction studies have been unable to distinguish any magnetic Bragg peaks. In this article, we present the results of a comprehensive investigation of the low-temperature phase in symmetry-broken polycrystalline Ce\textsubscript{2}O\textsubscript{3} using a combination of magnetic susceptibility, heat capacity, neutron diffraction, triple-axis and time-of-flight (TOF) inelastic neutron scattering (INS), and muon spin rotation ($\mu$SR). Our measurements and subsequent analysis confirm that the transition at T\textsubscript{N} can be associated with the ordering of moments on the Ce\textsuperscript{3+} site. Both a spontaneous magnetic order observed with $\mu$SR and a dispersive spin-wave spectrum observed with inelastic neutron scattering suggest a model wherein planar dipoles order antiferromagnetically. Notable inconsistencies between $\mu$SR and neutron scattering data within the dipole picture provide strong evidence for the ordering of higher-order moments.

\end{abstract}

% insert suggested keywords - APS authors don't need to do this
%\keywords{}

%\maketitle must follow title, authors, abstract, and keywords
\maketitle

\section{Introduction}
Cerium-based compounds are an interesting subset of materials that possess strongly correlated 4f electrons presenting both localized and itinerant behavior \cite{Johansson1974,Smith1983}. It is this property of the electrons that lends itself to a broad display of exotic phenomena in these materials, fueling much of their interest. Among these phenomena are strongly renormalized magnetic moments \cite{Ravot1980, Lawrence1982}, multipolar magnetic order \cite{Effantin1985,Kuwahara2009}, the Kondo effect \cite{Kondo1964}, heavy-fermion behavior \cite{Andres1975}, and heavy-fermion superconductivity \cite{Steglich1979}. Additional emergent phenomena have been reported in certain Ce-based systems, including observation of composite `vibronic' excitations arising from mixed phonon and crystal field excitations \cite{Thalmeier1982, Loewenhaupt2003, Adroja2012, Cermak2019}.

\begin{figure}
\includegraphics[width=\linewidth, trim = {0 0cm 0cm 0cm},clip]{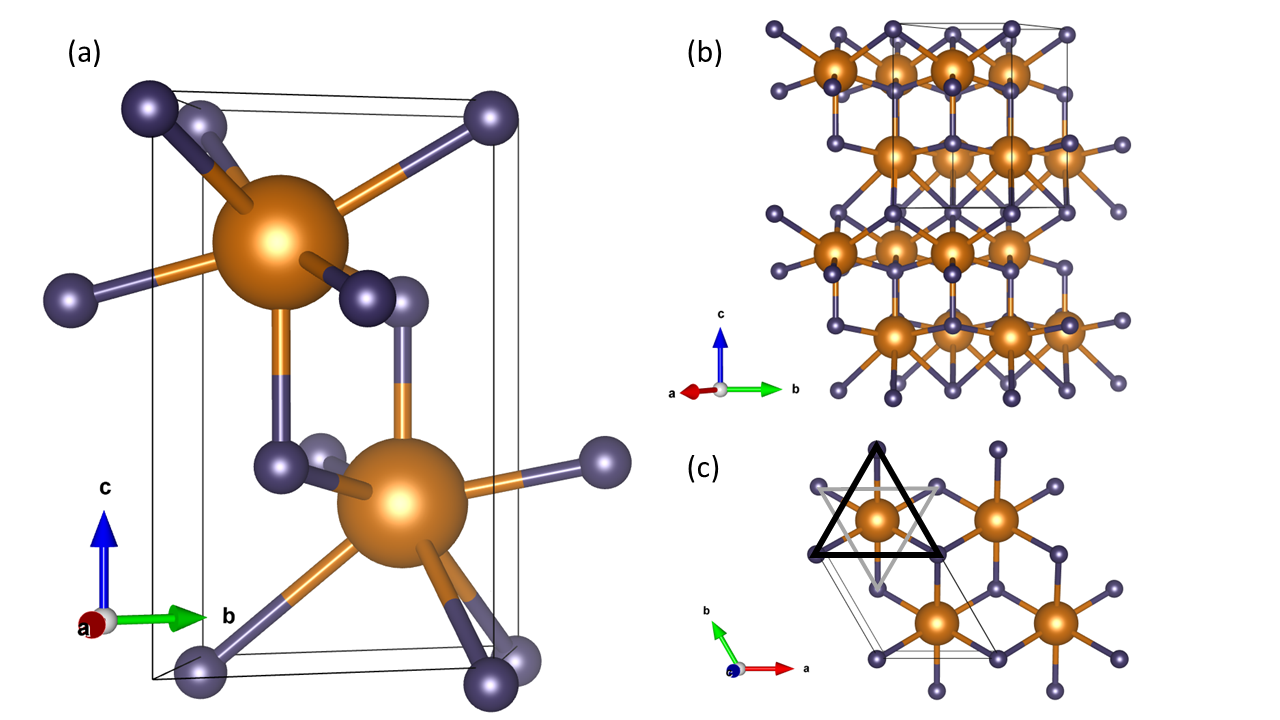}%
\caption{Crystal structure of Ce\textsubscript{2}O\textsubscript{3}.(a) Unit cell representation of structure. Orange spheres represent Ce atoms and blue spheres represent oxygen atoms. (b) Cross-section view illustrating alternating layers of O\textsuperscript{2-} and Ce\textsuperscript{3+} atoms. (c) View along the c-axis showing ab-planes from above. Oxygen atoms connected by black triangle are above the Ce-plane while those connected by grey triangle are below the Ce-plane.}
\label{fig:crystal}
\end{figure}

Here, we investigate the metastable oxide, Ce\textsubscript{2}O\textsubscript{3}. In this material the Ce\textsuperscript{3+} ion has C\textsubscript{3v} symmetry with 7-fold coordination to O\textsuperscript{2-} ligands (Fig.~\ref{fig:crystal}). The structure consists of bi-layer triangular nets of Ce atoms sandwiched by triangular nets of oxygen atoms, which crystallize in a trigonal $P\overline{3}m1$ phase. Lattice parameters are a = 3.88 {\AA} and  c = 6.05 {\AA} \cite{Barnighausen1985}, making the in-plane Ce-Ce distance 3.88 {\AA}, while Ce-Ce distances along c are slightly smaller at 3.74 {\AA} (intercell distance) and 3.80 {\AA} (intracell distance). Previous experiments examining the magnetic susceptibility and heat capacity of Ce\textsubscript{2}O\textsubscript{3} have identified a magnetic transition near T\textsubscript{N} = 6.2 K \cite{Justice1969, Kolodiazhnyi2018}, assumed to be associated with the onset of antiferromagnetic order. Additional investigations have reported unusual behaviors setting in near this transition. These include reports of a giant magnetodielectric effect near the ordering transition temperature, despite the lack of an obvious structural distortion \cite{Kolodiazhnyi2018}. Further, a Raman scattering study reported the emergence of mixed vibronic modes in Ce\textsubscript{2}O\textsubscript{3}  as the system was cooled below T\textsubscript{N} \cite{Sethi2019}, implying strong mixing between lattice and magnetic degrees of freedom.

While these experimental signatures are quite suggestive of a magnetic ordering transition in Ce\textsubscript{2}O\textsubscript{3}, conclusive evidence for the existence of such order has not yet been documented. Multiple investigations using neutron diffraction have failed to reveal any signature of magnetic Bragg peaks below T\textsubscript{N}, evoking comparison to the hidden order system URu\textsubscript{2}Si\textsubscript{2} \cite{Pinto1982,Kolodiazhnyi2018}. In one study, neutron diffraction revealed a realignment of crystallites in a powder sample with the application of a magnetic field, which the authors argued was consistent with an easy-plane anisotropy for ordered antiferromagnetic moments \cite{Kolodiazhnyi2018}. This was consistent with conclusions drawn from a modified crystal-field theory analysis in the same paper. The authors further used the observation of a linear magnetocapacitance and the lack of splitting between zero-field-cooled (ZFC) and field-cooled (FC) temperature curves in magnetization to argue for the existence of a low-temperature collinear antiferromagnetic phase with a \textbf{k}=0 propagation vector \cite{Kolodiazhnyi2018}. Another ordering picture was put forth in a recent study of the isostructural analogue Nd\textsubscript{2}O\textsubscript{3}, wherein magnetic Bragg peaks observed with neutron diffraction were indexed to a \textbf{k}=(1/2,0,1/2) propagation vector \cite{Sala2018}. At this time, the exact nature of magnetic order in Ce\textsubscript{2}O\textsubscript{3} remains an open question.

As the exotic behavior seen in this material is inextricably linked to the putative order, a more thorough examination is essential. In this article, we present the results of a multifaceted approach to probe the low temperature phase in Ce\textsubscript{2}O\textsubscript{3}. We investigate polycrystalline samples with multiple techniques, including neutron diffraction, time-of-flight (TOF), triple-axis inelastic neutron scattering (INS), and muon spin rotation ($\mu$SR). Our TOF data allowed us to characterize the crystal electric field (CEF) spectra associated with the Ce$^{3+}$ cations, the analysis of which confirms the conclusion of planar cerium spins. TOF and triple-axis data demonstrate a shifting and broadening of the CEF line at E=27 meV, in line with previous observations from Raman scattering of emergent vibronic modes. Significantly, TOF also revealed the existence of dispersive magnon excitations at lowest temperatures, where a precessive signal in $\mu$SR spectra implies the existence of magnetic order. These observations constitute the first direct evidence of long-range magnetic order in this material.

\section{Experimental Methods}
Polycrystalline samples of Ce\textsubscript{2}O\textsubscript{3} were prepared following the procedure laid out in Ref.~\onlinecite{Kolodiazhnyi2018} and characterized first using heat capacity and magnetization. Heat capacity measurements were taken using a Quantum Design PPMS Dynacool instrument, and magnetization measurements were performed with a Quantum Design MPMS3 Vibrating Sample Magnetometer. The PPMS and MPMS systems are both housed in the Stewart Blusson Quantum Matter Institute at the University of British Columbia (UBC).

Phase purity and characterization of the structure were determined using both x-ray diffraction (XRD) and neutron powder diffraction (NPD) measurements. Powder XRD was measured on a Bruker D8 Advance instruments both at UBC and in the Illinois Materials Research Laboratory using a copper K-$\alpha$ source and a Johansson monochromater that blocks any K-$\alpha_2$. NPD measurements were taken using the POWGEN high-resolution neutron powder diffractometer at the Spallation Neutron Source of Oak Ridge National Laboratory and the BT-1 the high-resolution neutron powder diffractometer at the NIST Center for Neutron Research (NCNR). Measurements at POWGEN were performed using an Orange cryostat at T = 2 K, 20 K, and 294 K with central wavelength 0.8 \r{A} and at T = 2 K, 3 K, 4 K, 5 K, 6 K, 7 K, and 10 K using central wavelength 2.67 \r{A}. Measurements at BT-1 were taken at 2 K and 20 K using a Cu (311) monochromator and a closed cycle refrigerator (CCR). Structural refinements were carried out using the GSAS-II software \cite{Toby2013}.

Inelastic neutron scattering measurements were carried out using both the TOF direct-geometry chopper spectrometer Merlin at the ISIS Muon and Neutron Source \cite{Bewley2006, Bewley2009} and the triple-axis spectrometer (TAS) BT-7 at the NCNR \cite{Lynn2012}. TOF measurements on Merlin used the Gd chopper at 500 Hz to obtain neutrons of incident energy E\textsubscript{i} = 16 meV, 23 meV, 66 meV, and 160 meV and at 350 Hz for neutrons of incident energy E\textsubscript{i} = 12 meV, 20 meV, 38 meV, and 100 meV. Samples on Merlin were measured at 1.8 K, 10 K, and 140 K. Resultant scans were normalized using comparable measurements of a vanadium standard sample. TAS measurements were taken with incident neutrons of energy 14.7 meV, selected with a pyrolytic graphite monochromator, and collimations of open-80'-80'-open. The sample was loaded in a He gas environment into an aluminum can with an indium seal. Measurements carried out at temperatures T= 2.6 K, 4 K, 5 K, 5.5 K, 6 K, 7 K, 8 K, and 15 K using a CCR.

$\mu$SR measurements were performed at TRIUMF using the LAMPF spectrometer in a low-background configuration on the M20 beamline in a helium flow cryostat. Zero-field (ZF) $\mu$SR measurements were then taken at various temperatures above and below the magnetic transition temperature. Longitudinal-field measurements were also taken at base temperature (T = 1.9 K) and in applied fields ranging from H = 100 Oe to 2000 Oe.

\begin{figure}[b]
\includegraphics[width=\linewidth, trim = {0 0cm 23cm 2.2cm},clip]{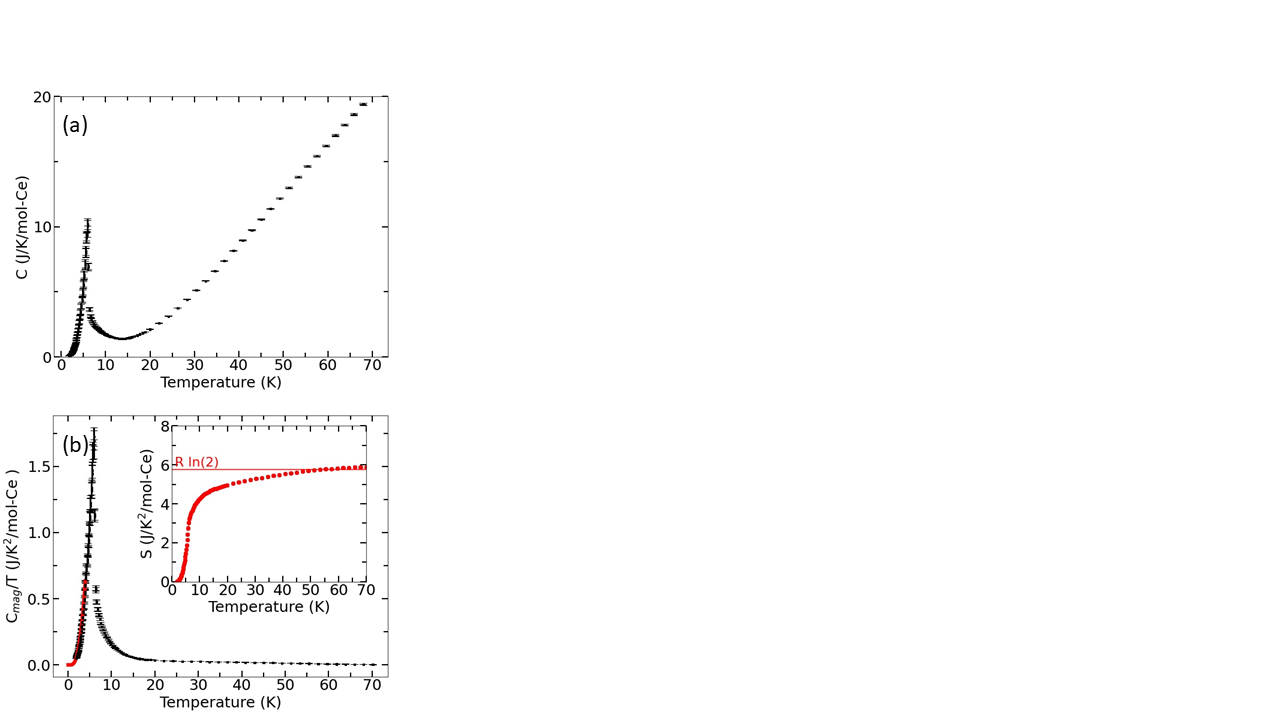}%
\caption{(a) Heat capacity of Ce\textsubscript{2}O\textsubscript{3} vs. temperature (raw data). (b) Magnetic contribution to heat capacity. Solid red line indicates fit to low-temperature data. Inset: Entropy from magnetic contribution to heat capacity.}
\label{fig:heatcap}
\end{figure}

\section{Heat Capacity}
Heat capacity measurements were taken from 1.8 K to 70 K, resulting in the curve shown in Figure~\ref{fig:heatcap}(a). The overall temperature dependence was consistent with measurements published elsewhere \cite{Kolodiazhnyi2018} and contained contributions from magnons, phonons and CEF excitations. To isolate the magnetic contribution, the CEF and lattice contributions were estimated and subtracted. The lattice contribution was estimated from measurements on the isomorphic La\textsubscript{2}O\textsubscript{3} \cite{Levels1963} and polynomial extrapolations to low temperature, as outlined in Ref.~\onlinecite{Kolodiazhnyi2018}. The CEF contribution can be computed directly from the three Kramers doublets observed with TOF neutron inelastic scattering, as described later in this paper. The heat capacity arising from these modes is given by the expression:
\begin{equation}
	C_{v}= \frac{1}{k_{B}T^{2}}(\frac{1}{Z}\sum_{n} E_{n}^{2}exp(-\beta E_{n}) - [\frac{1}{Z}\sum_{n} E_{n}exp(-\beta E_{n})]^{2})
\end{equation}
where $k\textsubscript{B}$ is the Boltzmann constant, $Z$ is the partition function, and $E_{n}$ are the CEF energy levels. A small contribution from a CeO\textsubscript{2} impurity identified with x-ray diffraction was also subtracted \cite{Westrum1960}.

The resulting magnetic contribution is shown in Figure~\ref{fig:heatcap}(b) and reveals most prominently a lambda-anomaly at T = 6.2 K, indicating a second-order phase transition. Integrating with temperature, we find that the entropy (inset of Fig.~\ref{fig:heatcap}(b)) approaches $Rln2$, as expected if we associate this transition with the ordering of effective spin degrees-of-freedom within the ground state doublet.

The heat capacity data at temperatures below 4 K fit well to an expression expected for antiferromagnetic spin wave materials \cite{Lashley2008}:
\begin{equation}
	C_{AF}={B_{AF}}{T}^{n}exp(-\Delta_{AF}/T)
\end{equation}
where $n$ was empirically fit to 3.14(14). $\Delta_{AF}$ represents the magnitude of the spin gap and fit to a value of 3.08(46) K while B$_{AF}$ fit to a value of 0.14(4).

\section{Magnetic Susceptibility}
\begin{figure}
\includegraphics[width=\linewidth, trim = {0.5cm 1.5cm 21cm 0cm},clip]{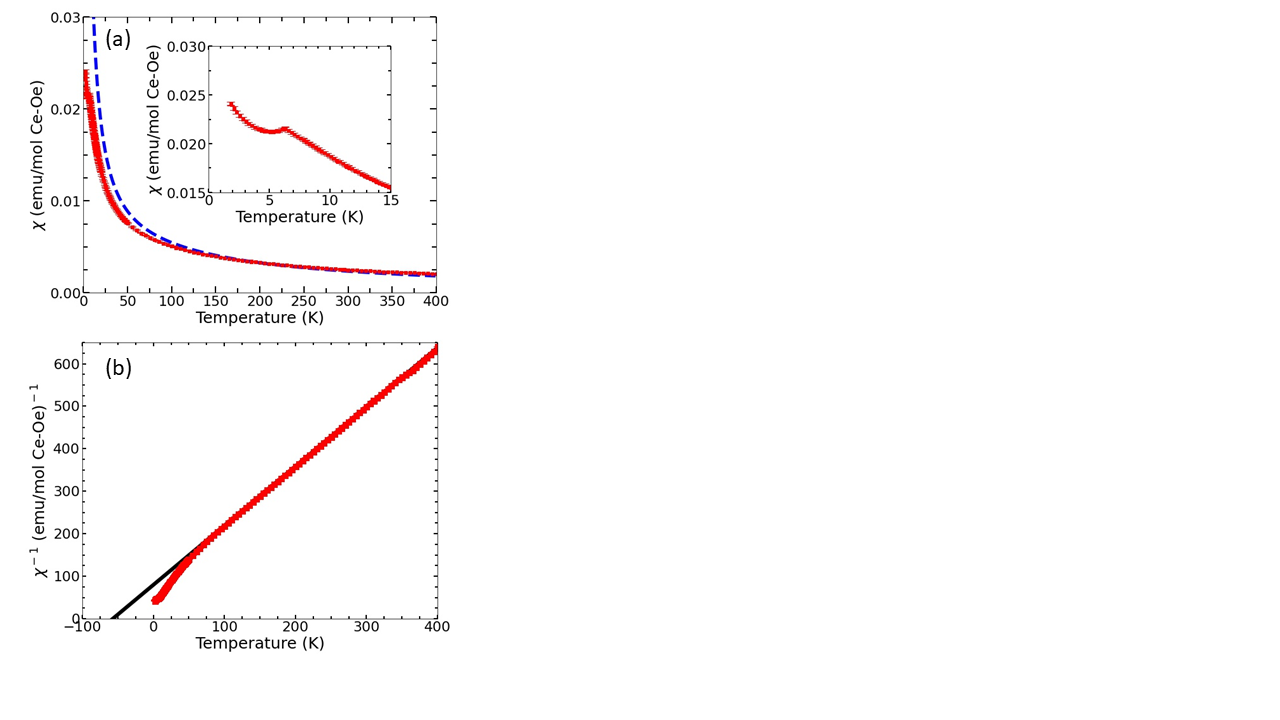}%
\caption{(a) Magnetic susceptibility of Ce\textsubscript{2}O\textsubscript{3} vs temperature (red squares). Blue dashed line corresponds to expected susceptibility with calculation as described in the text. Inset: Susceptibility near T\textsubscript{N}. (b) Inverse susceptibility with solid line representing fit to data as discussed in text.}
\label{fig:susc}
\end{figure}

Magnetic susceptibility measurements were performed in the temperature range T = 1.8 K - 400 K with an applied field H = 1000 Oe. Results taken using both ZFC and FC protocols are plotted in Figure~\ref{fig:susc}(a). High temperature behavior was well-described by a Curie-Weiss function, $\chi$=$\chi_0$+C/(T-$\theta$), as demonstrated by the plot of inverse susceptibility shown in Figure~\ref{fig:susc}(b). Fits of the data in the temperature region 100 K \textless T \textless 400 K to this function yielded an effective moment size $\mu_{eff}$ = 2.40(4) $\mu$\textsubscript{B} and Weiss constant $\theta$ = -57.0(5) K. This moment size is slightly smaller than the free-ion value of 2.54$\mu$\textsubscript{B} for Ce\textsuperscript{3+} with J=5/2. The Weiss constant is significantly larger than the ordering of T\textsubscript{N} = 6.2 K, indicating a moderate level of frustration. This may be a consequence of competition between the three nearest-neighbor Ce\textsuperscript{3+} interactions described above, as seen in Nd\textsubscript{2}O\textsubscript{3} \cite{Sala2018}.
The expected susceptibility was calculated using the eigenvectors and eigenenergies derived from the CEF analysis utilizing the MANTID package \cite{Arnold2014}. The data are reasonably well-described by the simulation at high temperatures, but diverge at lower temperatures, likely the result of magnetic interactions which are not accounted for in this model.
At the ordering temperature, the susceptibility reaches a cusp suddenly, consistent with a sharp onset of antiferromagnetic correlations. No discernible splitting is seen between ZFC and FC curves, arguing against any weak ferromagnetic moment or significant disorder. %The Curie-like tail at lowest temperatures may be associated with the known fraction of $CeO_2$, which can exhibit paramagnetic behavior \cite{Lipp2016}.

\section{Powder Diffraction}
\begin{table*}
\caption{Results of x-ray and neutron refinements on Ce\textsubscript{2}O\textsubscript{3} in space group $P\overline{3}m1$}. Uncertainties are statistical in origin and represent one standard deviation.
\begin{ruledtabular}
\begin{tabular}{l|cccccc}
Instrument & \multicolumn{2}{c}{Lattice Parameters} & \multicolumn{3}{c}{Atom Positions (Fractional Coordinates)} & Rw \tabularnewline
 & a (\AA) & c (\AA) & Ce  & O1 & O2 \tabularnewline
 \hline
Bruker D8 at 293 K & 3.8906(2)& 6.0658(1) & 1/3, 2/3, 0.2478(19)&0, 0, 0&1/3, 2/3, 0.661(10) & 5.158 \tabularnewline
POWGEN at 300 K & 3.8917(1)& 6.0585(1) & 1/3, 2/3, 0.2451(2)&0, 0, 0&1/3, 2/3, 0.6474(1) & 2.054\tabularnewline
POWGEN at 25 & 3.8833(1)& 6.0459(1)& 1/3, 2/3, 0.2448(1)&0, 0, 0&1/3, 2/3, 0.6472(1) & 2.412 \tabularnewline
POWGEN at 2 K & 3.8835(1)& 6.0463(1) & 1/3, 2/3, 0.2451(1)&0, 0, 0&1/3, 2/3, 0.6474(1) & 0.948\tabularnewline
BT-1 at 20 K & 3.8833(1) &6.0526(1) & 1/3, 2/3, 0.2437(2)&0, 0, 0&1/3, 2/3, 0.6486(2) & 6.73 \tabularnewline
BT-1 at 2 K & 3.8825(1) &6.0511(1) & 1/3, 2/3, 0.2440(2)&0, 0, 0&1/3, 2/3, 0.6482(2) & 6.794 \tabularnewline
\end{tabular}\end{ruledtabular}
\label{flo:refinements}
\end{table*}

As detailed above, diffraction measurements were performed using both XRD at room temperature and NPD at multiple lower temperatures. We did not find, in the diffraction patterns, any sign of impurities other than a persistent fraction of $CeO_2$, which can form from metastable $Ce_2O_3$ at a rate set by storage conditions. $CeO_2$ has no identifiable magnetic transitions, and this impurity did not affect the conclusions of this paper. Nonetheless, XRD was performed at room temperature using lab-based sources before every inelastic neutron, muon or thermodynamic measurement to ascertain the impurity fraction. The POWGEN TOF NPD instrument at the SNS has an extended range of coverage in reciprocal space (Q) and was used to determine the structure of $Ce_2O_3$ with a high degree of precision over an extended temperature region. Additional NPD measurements were performed at lowest temperatures using the BT-1 instrument at the NCNR to more carefully look for magnetic Bragg peaks at low scattering angles.

\begin{figure}
\includegraphics[width=\linewidth, trim = {1.2cm 0.2cm 20cm 0cm},clip]{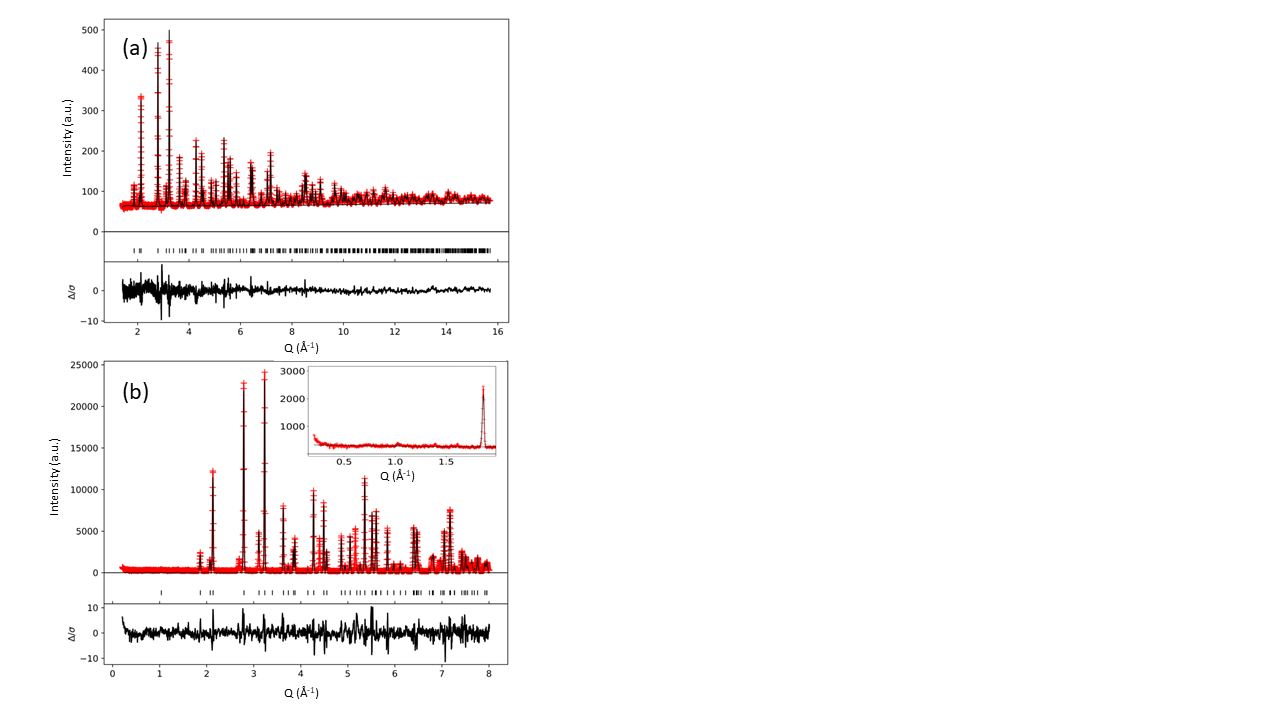}%
\caption{Neutron powder diffraction data from (a) POWGEN measured at T = 2 K with center wavelength $\lambda$ =  0.8 $\AA^{-1}$ and (b) BT-1 measured at 2K with wavelength $\lambda$ = 1.54 $\AA^{-1}$. Solid black lines depict fits to the $P\overline{3}m1$ structural model appropriate for Ce\textsubscript{2}O\textsubscript{3}. Additional peaks in (b) arise from Al can. Inset shows enlarged region at low Q where magnetic Bragg peaks are expected.}
\label{fig:NPD}
\end{figure}

Example patterns obtained using the POWGEN and BT-1 instruments can be found in Fig.~\ref{fig:NPD}. It was found that all diffraction data sets are well described by the previously reported $P\overline{3}m1$ structure for $Ce_2O_3$, and the results of Rietveld refinements on several typical data sets are given in Table~\ref{flo:refinements}. Sample parameters are found to be consistent both with the literature \cite{Kolodiazhnyi2018} and between fits performed on different instruments at comparable temperatures. Notably, the quality of fits to $P\overline{3}m1$ also did not worsen considerably for NPD measurements performed at the lowest temperatures at either BT-1 or POWGEN, which limits the size of any structural distortion which might exist at the ordering transition at T\textsubscript{N} = 6.2 K. As with previous attempts to investigate Ce\textsubscript{2}O\textsubscript{3} with NPD \cite{Kolodiazhnyi2018, Pinto1982}, our data revealed no evidence for magnetic Bragg peaks below T\textsubscript{N}. Assuming a traditional $Ce^{3+}$ form factor, this allows us to simulate the magnetic contribution to the powder diffraction pattern and determine an upper limit on the ordered magnetic moment. In doing so we calculate a maximum magnetic moment of $\mu$ = 0.32(11)$\mu_B$ associated with a predicted \textbf{k}=0 ordered state \cite{Kolodiazhnyi2018} or $\mu$ = 0.026(12)$\mu_B$ for a state with \text{k}=(1/2 0 1/2) as seen recently in Nd\textsubscript{2}O\textsubscript{3} \cite{Sala2018}.

\section{TOF Inelastic Neutron Scattering}

\begin{figure*}
	\includegraphics[width=\linewidth, trim = {0cm 3cm 0cm 0cm},clip]{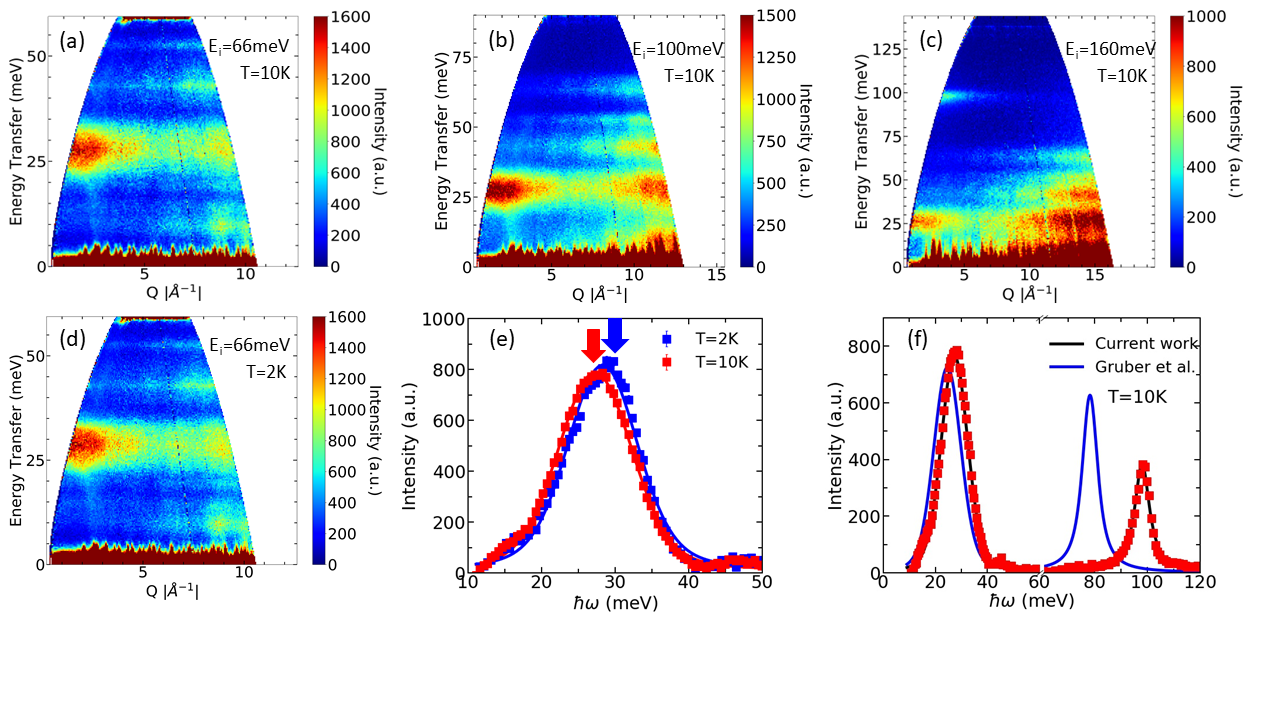}%
	 	 \caption{Time-of-flight inelastic neutron scattering measurements on Ce\textsubscript{2}O\textsubscript{3} with neutrons at E\textsubscript{i} = 66 meV (a,d), 100 meV (b) and 160 meV (c). Constant-Q cuts of data with scaling as described in text (e). Constant-Q cuts of data with fits represented by solid black line (f). Solid blue line depicts results from crystal field analysis performed by Gruber et al. \cite{Gruber2002}}. %
	\label{fig:TOF1}
\end{figure*}

\begin{figure}
%	 \centering\par\vspace{-2\baselineskip}
%	 \subfloat{{\includegraphics[width=6cm, trim = {0 1.6cm 0 2cm}]{magnon_peak.jpg}}}%
%	 \subfloat{{\includegraphics[width=6cm, trim = {1cm 1cm 1cm 4cm}]{MERLIN_magnon_final.png}}}% \\
%	 \subfloat{{\includegraphics[width=6cm, trim = {0 1cm 0 1cm}]{SPINW_FINAL.jpg}}}%
	\includegraphics[width=\linewidth, trim = {0cm 5.5cm 18cm 0cm},clip]{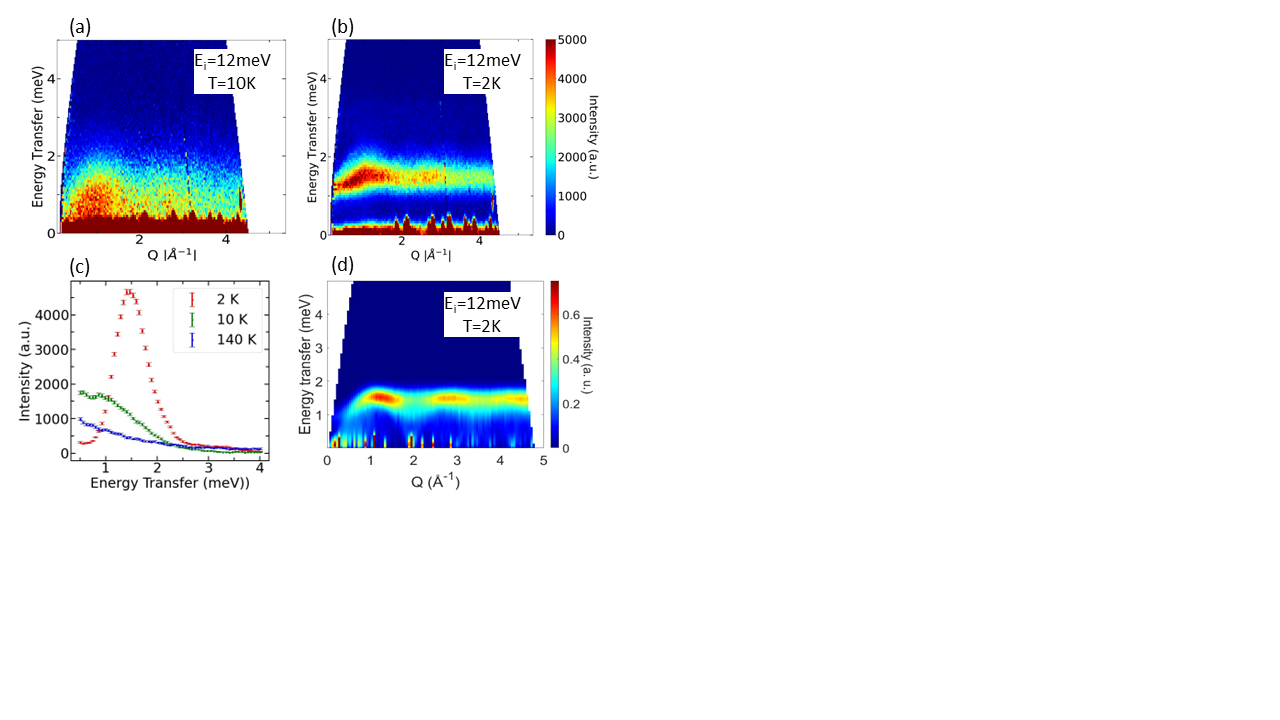}%
	 \caption{Inelastic neutron scattering with E\textsubscript{i}=12 meV at temperature above (a) and below transition (b). Constant-Q cuts of INS data where $\mid$Q$\mid$=0.8-1.2 $\AA^{-1}$ (d) Simulation of spin waves as described in text.} %
	 \label{fig:TOF3}
\end{figure}

The main results of our TOF inelastic neutron scattering measurements are summarized in Figs.~\ref{fig:TOF1} and \ref{fig:TOF3}. In Figs.~\ref{fig:TOF1} (a), (b) and (c), we show the normalized scattering data taken using neutrons with incident energies $E_i$ = 66 meV, 100 meV, and 160 meV, respectively, and temperature T = 10 K. Consistent across all spectra, several phonon excitations with a $I(Q) \propto$ Q\textsuperscript{2} momentum transfer dependence are apparent at energies $E\sim$ 9.4, 13.5, 19.4, 27, 43.5, 52.2, and 63 meV. These are equivalent to the phonon excitations reported for the structural analogue La\textsubscript{2}O\textsubscript{3} \cite{Petretto2018,Miranda}, and are as expected for this material. In addition, magnetic excitations were observed near $E\sim$ 27 and 98 meV, which we associate with CEF excitations of the Ce$^{3+}$ cations.

Notably, we see scattering from both phonons and CEF modes at the singular energy $E$ = 27 meV. This is in line with the Raman scattering measurements of Sethi \textit{et al.} \cite{Sethi2019}, where emergent vibronic excitations were also seen at temperatures below $T = T_N$. The current measurements did not have sufficient energy resolution to confirm the existence of mode splitting, but the data at the lowest temperatures did reveal a shift in spectral weight consistent with this previous study \cite{Sethi2019}. We demonstrate this shift in Figs.~\ref{fig:TOF1} (d) and (e) and explore it in greater detail using neutron triple-axis spectroscopy, discussed in a later section.

\begin{table*}[t]
\caption{Calculated wavefunctions for Ce\textsubscript{2}O\textsubscript{3} determined using fits to the time-of-flight inelastic neutron scattering data. Crystal field parameters are also included in the first line of the table.}
\begin{ruledtabular}
\begin{tabular}{l|cccccc}
        & $B_{2}^0 $=4.26 meV   & & $B_{4}^0 $ = 0.0339 meV  &    & $B_{4}^3 $ =-3.3516 meV    \tabularnewline
 \hline
E (meV) &$| -\frac{5}{2}\rangle$ & $| -\frac{3}{2}\rangle$ & $| -\frac{1}{2}\rangle$ & $| \frac{1}{2}\rangle$ & $| \frac{3}{2}\rangle$ & $| \frac{5}{2}\rangle$ \tabularnewline
 \hline
0.000 & 0.3455 & 0.0 & 0.0 & -0.9384 & 0.0 & 0.0 \tabularnewline
0.000 & 0.0 & 0.0 & 0.9384 & 0.0 & 0.0 & 0.3455 \tabularnewline
27.107 & 0.0 & -1.0 & 0.0 & 0.0 & 0.0 & 0.0 \tabularnewline
27.107 & 0.0 & 0.0 & 0.0 & 0.0 & -1.0 & 0.0 \tabularnewline
98.063 & -0.9384 & 0.0 & 0.0 & -0.3455 & 0.0 & 0.0 \tabularnewline
98.063 & 0.0 & 0.0 & -0.3455 & 0.0 & 0.0 & 0.9384 \tabularnewline
\end{tabular}\end{ruledtabular}
\label{flo:Eigenvectors}
\end{table*}

To extract information about the nature of the local Ce$^{3+}$ moments, we fit the intensity profiles of the two magnetic modes observed in our multiple data sets at $T$ = 10 K to a simple crystal field model. Magnetic scattering intensity was first isolated by taking constant-$Q$ cuts of normalized scattering data, integrating over the range $Q$ = [3.5,6] \AA$^{-1}$, and subtracting the scattering contribution from phonon scattering as determined by scaling data in the range $Q$ = [9.6,11.6] \AA$^{-1}$. The magnetic scattering near $E$ = 98 meV was characterized solely using neutrons with $E_i$ = 160 meV neutrons, whereas the magnetic scattering near $E$ = 27 meV was characterized through simultaneous consideration of $E_i$ = 100 meV and the $E_i$ = 66 meV data sets. The $E_i$ = 160 meV data was not considered when characterizing the $E$ = 27 meV region due to a high level of quasi-elastic scattering observed in the lower-resolution measurement at energies up to $E$ = 35 meV,

Our diffraction results showed that the Ce\textsuperscript{3+} atoms possess C\textsubscript{3v} point group symmetry, which implies non-zero values for $B_2^0$, $B_4^0$, and $B_4^3$ CEF parameters. As a first estimate of these parameters, we performed a simple point-charge model calculation, and then refined them further through consideration of the magnetic scattering data. Fits assumed a pseudo-Voigt energy profile for the CEF modes with varying widths and energies and intensities calculated by the software PyCrystalField \cite{Scheie2020}.
Best fit crystal field parameters are presented alongside the energy eigenvalues and wave functions in Table \ref{flo:Eigenvectors}. The quality of the fit is demonstrated in Fig.~\ref{fig:TOF1}, which compares the predictions of the current work to those of the thermodynamic analysis of Gruber \textit{et al.} \cite{Gruber2002}.

This CEF analysis reveals a ground state Kramer's doublet, as expected for a Ce$^{3+}$ cation, with two doublets removed to higher energies. Using the ground state wave functions and Equations \ref{eq:4} and \ref{eq:5}, we are able to calculate g\textsubscript{$\perp$}=2.26 and g\textsubscript{$\parallel$}=0.24:
\begin{equation}\label{eq:4}
	 g\textsubscript{$\perp$}=2g_{J}\mel{\Psi_{\downarrow}}{J\textsuperscript{+}}{\Psi_{\uparrow}}
\end{equation}
\begin{equation}\label{eq:5}
	 g\textsubscript{$\parallel$}=2g_{J}\mel{\Psi_{\uparrow}}{J_z}{\Psi_{\uparrow}}
\end{equation}

These values imply a total dipole moment of 1.6 $\mu_B$ with a distinct easy-plane spin anisotropy.
This conclusion about anisotropy is in agreement with the modified ligand field analysis of Kolodiazhnyi \textit{et al.} \cite{Kolodiazhnyi2018}, as well as studies investigating other materials housing Ce\textsuperscript{3+} ions in trigonal environments \cite{Banda2018}.

%the Stevens formalism \cite{Stevens1952}.
%\begin{equation}
%	H=\sum_{n,m} B_{n}^{m}O_{n}^{m}
% 	\phi(T) = \phi_{0}(1 - \frac{T}{T_{C})^{\beta}
%\end{equation}

Figure~\ref{fig:TOF3} shows the results of our measurements with $E_i$ = 12 meV neutrons, which allow for closer inspection of excitations near the elastic line. Upon lowering the temperature to $T$ = 10 K, these data reveal a significant build-up of paramagnetic scattering centered most strongly at $Q\approx$1 $\AA^{-1}$ (Fig.~\ref{fig:TOF3}(a)), which evolves into a dispersive spin-wave spectrum on cooling to $T$ = 2 K (Fig.~\ref{fig:TOF3}(b)). The magnon density of states is centered at $E \sim$ 1.5 meV and has bandwidth of $\Delta E \sim$ 1 meV  (Fig.~\ref{fig:TOF3}(c)), consistent in scale with an ordering transition at $T_N$ = 6.2 K.

For comparison, we show in Fig.~\ref{fig:TOF3}(d) a simulation of the expected scattering pattern from a powder of Ce\textsubscript{2}O\textsubscript{3} containing the $\mathbf{k} = 0$ collinear antiferromagnetism proposed by Kolodiazhnyi \textit{et al.} \cite{Kolodiazhnyi2018}. Powder-averaged simulations were performed using the SpinW software package \cite{Toth2015} and assumed dipole $Ce^{3+}$ moments, the g-tensor calculated above, an exchange coefficient J= 1 meV and a spin gap determined from fits of our heat capacity data ($\Delta$ =3.08 K). The simulation successfully reproduces several features of the observed scattering pattern, including the ridge of scattering near 1.5 meV and the positions in $Q$ of several local scattering maxima. The measured spin-wave scattering seems to fall away more rapidly than the dipole simulation, probably reflecting the significant multipolar character of the $Ce^{3+}$ moments which can be inferred from the wavefunctions for the electrons in the Kramers doublet determined above and the associated distribution of electron density in the unit cell. The apparent gap of ~$\sim$ 1 meV in the scattering data is understood as an effect of the powder density-of-states. Overall, the appearance of collective spin-wave excitations is a significant observation, which constitutes the strongest evidence to date of long-ranged spin order in Ce\textsubscript{2}O\textsubscript{3}.

\section{Triple-Axis Neutron Scattering}
\begin{figure*}
    \vspace{-0.5cm}
	\includegraphics[width=\linewidth, trim = {0cm 9cm 0cm 0cm},clip]{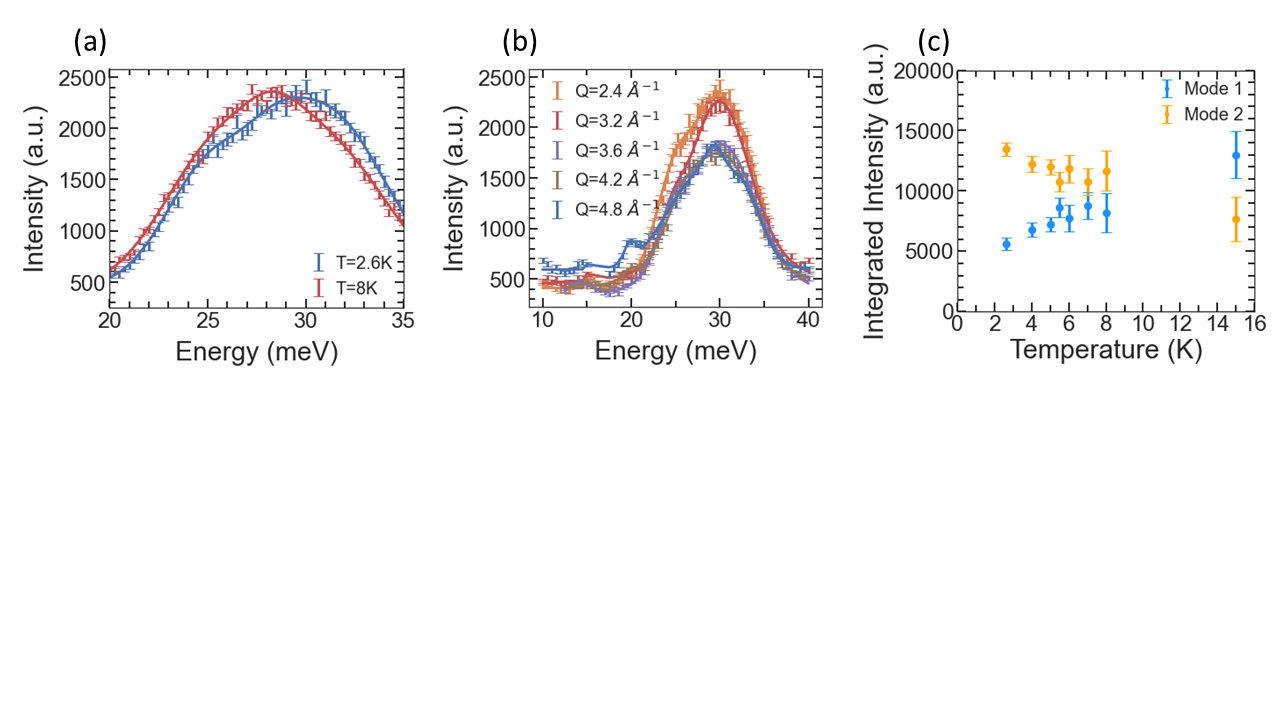}%
	 	 \caption{(a) Inelastic neutron scattering in the region around expected vibronic modes. Solid lines represent fits to data as described in text. (b) Constant-Q cuts at 2.6 K. (c) Temperature dependence of the integrated intensity of the peaks at approximately 26 meV (Mode 1) and 31 meV (Mode 2).} %
	\label{fig:INS1}
\end{figure*}

We followed our TOF inelastic studies with a more detailed study of the 27 meV mode with triple-axis neutron scattering. The underlying motivation was to confirm and expand upon recent Raman spectroscopy results which reported emergent vibronic excitations near this energy \cite{Sethi2019}. In particular, the major features observed in the Raman data were two new modes near a purported two-phonon band at 24 meV and the CEF excitation at 27 meV as the sample was cooled below T\textsubscript{N}. These emergent modes appeared at 22 meV and 31 meV and appeared to have phonon-like and electron-like behavior, respectively. Notably, a clear shift in spectral weight from the CEF mode into the 31 meV mode was also observed as temperature was decreased.

Our TOF data reveal a similar shift in spectral weight as the sample is cooled below T\textsubscript{N}, and this is reaffirmed by temperature-dependent TAS scans at Q = 2.4 \AA\textsuperscript{-1} (Figure \ref{fig:INS1}(a)). In examining constant-Q scans at base temperature, one can see that much of the scattering intensity is magnetic, as evidenced by the reduced intensity with increasing momentum transfer (Figure \ref{fig:INS1}(b)). Phonon scattering is relatively weak in this region of momentum space, but there may also be a phonon-like component, seen as a subtle inflection in scattering intensity at higher Q. This Q-dependence is similar to that seen in the TOF data for this region. Notably, the peak in this region remains broad, despite the increase in energy resolution of our TAS measurements over our TOF data. This is actually fairly consistent with the Raman scattering study, where the CEF modes were broad in energy and the phonon-like modes much sharper. The Q-range of our TAS measurements strongly suppressed the phonon scattering in favor of scattering from the magnetic modes. The fact that the width is much wider than instrument resolution implies that the magnetic modes have a finite lifetime, possibly due to scattering from phonons or delocalized electrons. We do not observe more than a single peak, but the observed mode has a distinct asymmetry. This might imply a profile consisting of at least two unresolved peaks, though best fits to these data are achieved using a sum of three Gaussian functions. Based on the Q-dependence of these data and the TOF data above, we associate these three peaks with one phonon at 28 meV and two magnetic excitations at 26 and 31 meV, respectively.

To characterize the temperature evolution of this scattering, the constant-Q scans were fit to the three Gaussian profile at all temperatures, and the best fits are shown as solid lines in Figs. \ref{fig:INS1}(a) and (b). As no significant temperature-dependence of the phonon mode at 28 meV was observed in the TOF data, fits for the phonon peak were made for the lowest temperatures only and then held constant for higher temperatures. Figure \ref{fig:INS1}(c) shows the fitted intensities of the two temperature-dependent modes, and implies an apparent transfer of intensity from the 26 meV mode (Mode 1) into the mode at 31 meV (Mode 2). This is consistent with the previous Raman study, though we do not see the additional peak at approximately 22 meV, likely due to its heavily phonon-like nature. This limitation and the large intrinsic width prevents us from commenting more explicitly on the presence or nature of possible vibronic modes, which were clearly observed in CeCuAl\textsubscript{3} \cite{Adroja2012}. A future investigation using polarized neutrons on single crystal samples is suggested to better separate phonon and magnetic contributions and track dispersion of these modes in three dimensional reciprocal space.

\section{Muon Spin Rotation}
Finally, we performed a series of $\mu$SR measurements to directly couple to the magnetic order parameter and glean further information about the nature of the low-temperature state. $\mu$SR is exquisitely sensitive to small magnetic fields inside materials and has a long, established history of detecting small moment spin order in f-electron systems. It is sensitive to both dipole and multipolar magnetism, and is largely unaffected by absorption, disorder, delocalization and other complicating factors which can obscure Bragg peaks in a neutron scattering experiment.

\begin{figure*}
	\includegraphics[width=\linewidth, trim = {0cm 0cm 0cm 0cm},clip]{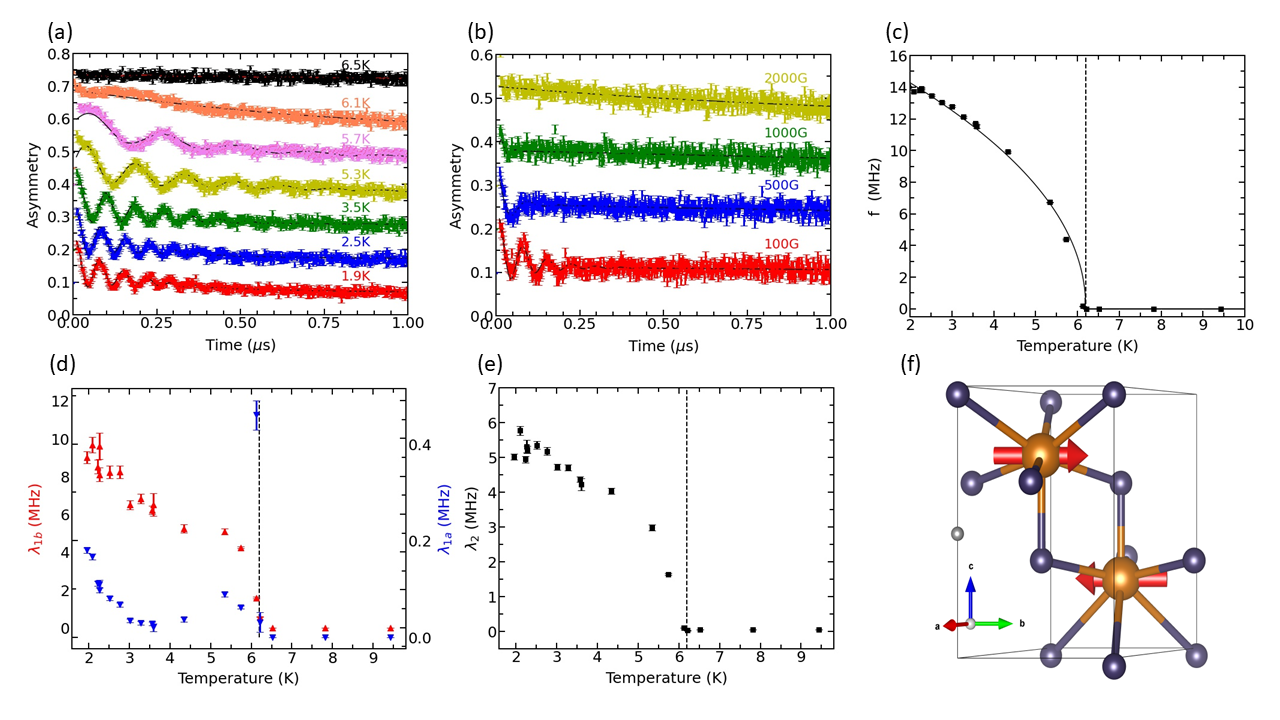}%
	 \caption{(a) Raw ZF-$\mu$SR spectra at various temperatures, where independent scans are offset by increments of 0.1 along the y-axis for data taken at temperatures between 1.9 K  and 5.7 K and by increments of 0.05 thereafter. Solid black lines represent fits using Eq.~\ref{eq:muon}. (b) Raw LF-$\mu$SR spectra taken at T = 1.9 K in several different applied fields. Solid black lines represent similar fits without constraints on the ratio between oscillating and non-oscillating fractions.  (c)-(e) Parameters extracted from fits of ZF-$\mu$SR spectra, plotted as a function of temperature. (f) A schematic of the atoms in a single unit cell, with the location of the predicted muon site shown as a grey sphere and the dipole spin direction in the \textbf{k} = 0 ordered state shown as red arrows.}
	\label{fig:muon1}
\end{figure*}

Figure \ref{fig:muon1} (a) shows spectra from several zero-field (ZF) $\mu$SR measurements taken at temperatures above and below the transition temperature at $T_N$ = 6.2 K. Well above $T_N$, the ZF spectra exhibit a weak Gaussian relaxation, as expected for a system of randomly oriented dipoles. Upon cooling, the most salient feature of the data is the emergence of clear oscillations below T$_N$ with a frequency that increases with decreasing temperature to a value of about $f$ = 14 MHz. This reflects the Larmor precession of muon spins around an emergent local field of H$_{local}$ = 0.1 T and provides direct and unambiguous evidence for long-ranged magnetic order. Longitudinal-field (LF-$\mu$SR) measurements performed at base temperature (T = 1.9 K) show that the spectra decouple in comparably sized fields (Figure [\ref{fig:muon1}] (b)), indicating that the order is static on the $\mu$SR timescale.

Zero-field spectra were fit to the muon spin depolarization function shown in Eq.~\ref{eq:muon}. In this equation, A\textsubscript{0} is the total muon asymmetry, which is a property of the spectrometer and determined by an independent calibration measurement performed in the paramagnetic state at T = 20 K.  A\textsubscript{1} is the fraction of the initial asymmetry representing the magnetically ordered volume and is itself separated into a 2/3 oscillating component and a 1/3 non-oscillating component, corresponding to mean field components perpendicular and parallel to the initial muon spin direction. A\textsubscript{2} and A\textsubscript{3} are non-oscillating components, with and without discernible time relaxation, respectively. Inclusion of these two non-oscillating components was determined empirically to be the minimal model required to explain the spectra at the lowest temperatures. All fits were performed under the constraint A\textsubscript{1} + A\textsubscript{2} + A\textsubscript{3} = 1, and the weight of the various fractions were found to be temperature independent well below the transition with approximate values A\textsubscript{1} = 0.51, A\textsubscript{2} = 0.34, and A\textsubscript{3} = 0.15. These values were held constant in fits close to the transition for stability. The size of the time-independent portion, A\textsubscript{3}, is consistent with the nonmagnetic CeO\textsubscript{2} impurity fraction determined using x-ray diffraction, whereas the non-oscillating component A\textsubscript{2} may represent either muons landing in a non-magnetic volume inside our Ce$_2$O$_3$ sample or muons residing at a symmetric location in the unit cell with zero mean field (where the likely muon stopping site is discussed below).

\begin{equation}\label{eq:muon}
\begin{aligned}
	A_{0}P(t) = A_{0}A_{1}(\frac{1}{3}exp^{-(\lambda_{1a} t)} + \frac{2}{3}\cos(2\pi f t + \phi)\exp^{-(\lambda_{1b} t)})\\
	+A_{2}A_{0}\exp^{-(\lambda_{2}t)}+A_{3}A_{0}
\end{aligned}
\end{equation}

As seen by the black lines in Fig.~\ref{fig:muon1}(a), fits using this equation were largely successful in describing the data at all temperatures, with the possible exception of near the transition where it misses at early times. Parameters extracted from these fits are shown in Figs.~\ref{fig:muon1}(c)-(e) and show clear and smooth evolution with temperature. The oscillation frequency in the ordered volume is plotted in Fig.~\ref{fig:muon1}(c) and shows a clear order-parameter temperature dependence, consistent with our interpretation of a second order phase transition. In this case thermal dependence of the oscillation frequency in the ordered phase can be described by:
\begin{equation}
	f(T) = f_{0}(1 -\frac{T}{T_{N}})^{\beta}
\end{equation}

Fits to this formula give a critical temperature of $T_N$ = 6.1 K, consistent with the value we determined from heat capacity, and the exponent $\beta$ fits to a value of 0.46, close to the value of $\beta=1/2$ expected from mean field theory. Fig.~\ref{fig:muon1}(d) shows the exponential relaxation rates associated with oscillating and non-oscillating components of the ordered fraction.  Here, $\lambda$\textsubscript{1b} represents the relaxation of the cosine component and should be associated with inhomogeneous broadening of the local field distribution at the muon site; relaxation here is proportional to the second moment of the distribution and has similar temperature dependence as the mean field, which is common. The parameter $\lambda$\textsubscript{1a} represents the relaxation of the non-oscillating component in the ordered volume and likely reflects a weak dynamic relaxation mechanism; it seems to diverge at the ordering temperature, consistent with critical fluctuations near a second order phase transition. Fig.~\ref{fig:muon1}(e) shows the exponential relaxation of the A\textsubscript{2} non-oscillatory fraction, which surprisingly has a similar temperature dependence and magnitude of frequency in the A\textsubscript{1} fraction. This implies that the local field distribution at the site of muons in the A\textsubscript{2} comprises stray fields from the ordered fraction. In fact, given the lack of net magnetization in the low-temperature region, the magnitude of the fields contributing to the A\textsubscript{2} strongly implies these muons reside within the ordered volume, but are located at symmetric positions with respect to the magnetic cations.

Beyond these details, it important to note that the size of the mean field implied by the precession frequency of A\textsubscript{1} muons is quite significant and provides great contrast to the tight upper limits on the possible ordered moment inferred from neutron diffraction data. To put this statement on more quantitative footing, we used the MuFinder software suite \cite{Huddart2021} to find the most likely muon stopping site via advanced DFT analysis. This was found to be at the fractional coordinates x = 0.000, y = 0.003, z = 0.458 in hexagonal coordinates as shown by the grey sphere in (Figure \ref{fig:muon1} (f)), favored over other stopping sites by an energy of greater than 300 meV. Assuming this site, we were able to calculate the mean dipole field likely felt by muons as a result of the ordered states suggested in the previous literature. For example, if we assumed the \textbf{k} = (1/2, 0, 1/2) ordered state with planar spins seen in isostructural Nd\textsubscript{2}O\textsubscript{3} \cite{Sala2018}, we infer from the oscillation frequency in the ZF-$\mu$SR data that the ordered moment would have to be approximately 0.4 $\mu$\textsubscript{B}. Assuming the \textbf{k} = 0 with easy-plane spins requires a moment size of 2.45-2.65 $\mu$\textsubscript{B}, depending on the angle of rotation in the planes. %The \textbf{k} = 0 with easy-axis spins would require a 1.3 $\mu$\textsubscript{B} moment.
Within the context of the dipole moment picture, neither of these calculated moment sizes are compatible with the restrictions imposed by our neutron diffraction data, which imply ordered moments smaller than 0.3 and 0.03 $\mu$\textsubscript{B} for the \textbf{k} = 0 and \textbf{k} = (1/2, 0, 1/2) structures, respectively. This paradox is easily resolved if one considers a potential role for higher order magnetic multipoles, the ordering of which would contribute to the local field distribution experienced by muons which would not contribute to a neutron diffraction pattern. The existence of multipolar order in Ce$_2$O$_3$ would put this material in line with a long list of other f-electron magnets \cite{Donni2000, Nakao2001}.

%This small moment size is not uncommon in Ce-based systems, where moments have been seen to differ dramatically from the free-ion value \cite{Ravot1980}.  Moreover, the appearance of quadrupolar moments has previously been linked to a sizable reduction in the moment \cite{Hirai2021}.

\section{Conclusions}
Altogether, the data presented in this article reveal several new properties and characteristics that resolve some of the outstanding mysteries surrounding the material Ce$_2$O$_3$ that are essential for a full understanding of the low-temperature magnetic properties. Most revealing, the emergence of clear oscillations in our $\mu$SR data demonstrating the development of magnetic order and the associated spin wave excitations observed in our TOF inelastic neutron scattering data provide the first direct confirmation of long-ranged magnetic order in this material below 6.2 K.  While the absence of observable magnetic Bragg peaks in our neutron powder diffraction patterns puts tight limits on the maximum size of ordered dipole moments, the presence of dispersive magnon excitations confirm that ordered moments must exist in some form. The success of our spin-wave simulations provides support for the \textbf{k} = 0 ordered state suggested by previous studies, with spins aligned in the easy planes. However, the interpretation of the $\mu$SR data utilizing this magnetic structure suggests an ordered moment size of approximately 2.5 $\mu$\textsubscript{B} within the dipole picture, impossible to reconcile with the conclusions from neutron diffraction. Similar inconsistencies have been seen previously in cases of hidden order emerging in f-electron magnets \cite{ Cameron2016, Shen2019a}. Perhaps the most notable example is that of URu\textsubscript{2}Si\textsubscript{2}, where the nature of the hidden order remains unresolved after decades of study \cite{Mydosh2020}. It seems an inescapable conclusion then that higher-rank multipolar moments may provide the answer to these seemingly contradictory observations. This further helps explain the more rapid than expected suppression of magnetic scattering with increasing Q. We note in passing that the involvement of higher-rank multipoles in the low temperature region may also help explain reports of a giant magnetodielectric effect and the appearance of vibronic excitations in Raman scattering data, since multipoles are tied to electron orbitals and thus distribution of electron charge density within the crystallographic unit cell. To illuminate this physics further, we suggest a combined theory and resonant x-ray scattering to determine the nature of the multipolar order in Ce$_2$O$_3$ and its coupling to other material properties.

\begin{acknowledgments}
We thank ISIS Facility for providing the beam time RB2000208 (Gregory MacDougall et al; (2021): Investigation of Vibronic Excitations in Ce\textsubscript{2}O\textsubscript{3}, STFC ISIS Neutron and Muon Source, https://doi.org/10.5286/ISIS.E.RB2000208-1). This work was supported by the the Natural Sciences and Engineering Research Council of Canada and the CIFAR Azrieli Global Scholars program. This research was undertaken thanks in part to funding from the Canada First Research Excellence Fund, Quantum Materials and Future Technologies Program. S.L.C. and J.E.S. were supported by the National Science Foundation under Grant No. NSF DMR 1800982. The identification of any commercial product or trade name does not imply endorsement or recommendation by the National Institute of Standards and Technology.
\end{acknowledgments}

% Create the reference section using BibTeX:
%\bibliography{Ce2O3_LRO}
%apsrev4-2.bst 2019-01-14 (MD) hand-edited version of apsrev4-1.bst
%Control: key (0)
%Control: author (8) initials jnrlst
%Control: editor formatted (1) identically to author
%Control: production of article title (0) allowed
%Control: page (0) single
%Control: year (1) truncated
%Control: production of eprint (0) enabled
%

\end{document}